\documentclass[12pt,preprint]{aastex}

\cpright{AAS}{2009}
\shorttitle{Mixed Helicity in Erupting Filaments}
\shortauthors{Muglach, Wang, \& Kliem}

\begin{document}

\title{Evidence For Mixed Helicity in Erupting Filaments}
\author{K. Muglach,\altaffilmark{1} Y.-M. Wang, and B. Kliem\altaffilmark{2,3}}
\affil{Space Science Division, Naval Research Laboratory, 
Washington, DC 20375-5352}
\email{muglach@nrl.navy.mil, yi.wang@nrl.navy.mil, bkliem@uni-potsdam.de}
\altaffiltext{1}{ARTEP, Inc., Ellicott City, MD.}
\altaffiltext{2}{MSSL, University College London, Holmbury St.~Mary, Dorking, 
Surrey, UK.}
\altaffiltext{3}{Institut f{\"u}r Physik und Astronomie, 
Universit{\"a}t Potsdam, Potsdam, Germany.}
\slugcomment{ApJ, accepted}

\begin{abstract}
Erupting filaments are sometimes observed to undergo a rotation about the 
vertical direction as they rise.  This rotation of the filament axis 
is generally interpreted as a conversion of twist into writhe in a 
kink-unstable magnetic flux rope.  Consistent with this interpretation, 
the rotation is usually found to be clockwise (as viewed from above) 
if the post-eruption arcade has right-handed helicity, but counterclockwise 
if it has left-handed helicity.  Here, we describe two non--active-region 
filament events recorded with the Extreme-Ultraviolet Imaging Telescope (EIT) 
on the {\it Solar and Heliospheric Observatory} ({\it SOHO}), in which the 
sense of rotation appears to be opposite to that expected from the helicity 
of the post-event arcade.  Based on these observations, we suggest that 
the rotation of the filament axis is in general determined by the 
net helicity of the erupting system, and that the axially aligned core of 
the filament can have the opposite helicity sign to the surrounding field.  
In most cases, the surrounding field provides the main contribution to 
the net helicity.  In the events reported here, however, the helicity 
associated with the filament ``barbs'' is opposite in sign to and dominates 
that of the overlying arcade.
\end{abstract}


\section{INTRODUCTION}
Hanle effect measurements (Bommier et al. 1994) have shown that 
quiescent (non--active-region) filaments have ``inverse'' magnetic polarity, 
meaning that their transverse field component is opposite in sign to that 
of the coronal loops crossing high over the photospheric neutral line.  
The usual interpretation of this result is that the filament material 
is supported in the dipped or concave-upward portions of a twisted flux rope 
(see, e.g., van Ballegooijen \& Martens 1989; Rust \& Kumar 1994; Low 1996; 
Aulanier \& D{\'e}moulin 1998; Amari et al. 1999; Lites 2005; Gibson \& 
Fan 2006).  For this topology, it is natural to assume that one sign of 
helicity prevails throughout the flux rope and its overlying arcade.  

In an alternative scenario, the filament is regarded as a sheared arcade 
whose intermediate legs or ``barbs'' are rooted in minority polarity 
on the ``wrong'' side of the neutral line (Martin 1998).  In that case, 
the helicity of the filament itself would be opposite in sign to that of 
the overlying coronal loops, contrary to most theoretical models.  
Possible support for this picture comes from the observation of 
continual flows along the spines and up and down the barbs of filaments 
(Zirker et al. 1998; Wang 1999; Kucera et al. 2003; Lin et al. 2003, 2005); 
such flows, which may be driven by chromospheric reconnection (Litvinenko \& 
Martin 1999; Martin et al. 2008), seem to obviate the need for 
static support within magnetic dips, one of the principal arguments 
in favor of a flux rope geometry.

Statistically (and independently of the flux rope/sheared arcade question), 
it is found that filaments in the northern (southern) hemisphere 
are predominantly ``dextral'' (``sinistral''), meaning that 
their axial fields point to the right (left) when viewed from the 
positive-polarity side of the photospheric neutral line (Martin 1998; 
Pevtsov et al. 2003; Yeates et al. 2007; Yeates \& Mackay 2009).  
Dextral (sinistral) filaments have right-bearing (left-bearing) barbs 
but underlie arcades with left-handed (right-handed) helicity.  
The helicity sign of the overlying arcade can be inferred from the 
skew of the arcade loops relative to the polarity inversion line (PIL).

Although the magnetic topology of a stable filament/prominence remains 
a subject of debate, it is almost universally accepted that an erupting 
filament and its surrounding field have the structure of a twisted flux rope.  
In a three-dimensional system, the pinching-off of pairs of stretched, 
opposite-polarity field lines beneath the rising filament will necessarily 
give rise to a flux rope; reconnections may occur among the legs/barbs 
of the filament, between the legs and the overlying arcade, and among the 
highly distended arcade field lines themselves.  As the bright post-eruption 
arcade expands outward from the PIL, the orientation of the reconnected loops 
(seen projected against the photosphere) will appear to rotate 
clockwise (counterclockwise) if the arcade has right-handed (left-handed) 
helicity; this apparent rotation is due to the tendency for the 
loops rooted farther from the PIL to cross it more nearly at right angles.

The axes of the erupting filaments themselves are sometimes observed 
to rotate about the vertical direction.  As viewed from above, this rotation 
is again usually clockwise if the post-event arcade has right-handed helicity, 
and counterclockwise if it has left-handed helicity (for examples, see 
Rust \& LaBonte 2005; Green et al. 2007; Wang et al. 2009).  The sense of 
rotation is consistent with the conversion of field-line twist into writhe 
(three-dimensional twisting of the axis itself), as predicted by 
MHD simulations of kink-unstable flux ropes (Linton et al. 1999; Kliem 
et al. 2004; T{\"o}r{\"o}k \& Kliem 2005) and of sheared arcades 
that have been converted into erupting flux ropes via reconnection (Lynch 
et al. 2009).  It is reasonable to assume that the net helicity of the 
erupting structure is the sum of the helicity of the original filament 
and that of the surrounding field with which it has become entrained by 
reconnection (see Berger 1998).  If the barb system and the overlying coronal 
field have opposite helicity signs, as suggested by Martin (1998), then in 
most events the main contribution to the helicity of the erupting structure 
must come from the surrounding field (compare also the discussion of 
Ruzmaikin et al. 2003).  However, the question arises as to whether 
the rotation ever occurs in the direction opposite to that implied by 
the helicity of the overlying arcade, thereby providing stronger support 
for the barb picture of Martin.

Here we describe two events observed with the 
Extreme-Ultraviolet Imaging Telescope (EIT; Delaboudini{\`e}re et al. 1995) 
on the {\it Solar and Heliospheric Observatory} ({\it SOHO}), in which 
the erupting quiescent filament appeared to rotate counterclockwise but the 
post-eruption arcade had right-handed helicity.

\section{EVENT OF 1999 SEPTEMBER 20}
The sequence of \ion{Fe}{12} 19.5~nm images in Figure~1 shows the eruption 
of a southern-hemisphere filament on 1999 September~20.  The dark filament 
in fact represents the western half of a longer structure that occupies 
a circular filament channel.  The axis of the erupting feature undergoes a 
continual counterclockwise rotation between $\sim$03:00 and $\sim$05:00~UT.  
In order to exhibit this rotation more clearly, we have traced the 
filament spine at successive times and superposed these traces in Figure~2, 
after shifting them in longitude to correct for the average photospheric 
rotation.  We note that the observed motions cannot be interpreted simply 
as a straightening of the initially highly curved filament as it rises, 
which would require both the northern and southern portions of the filament 
to move eastward; instead, the northern section moves eastward but the 
southern section moves westward, consistent with a counterclockwise rotation 
of the filament axis.  Subsequently (after 04:48~UT), the rotation 
``overshoots'' near both the northern and southern ends of the 
erupting filament, which thus kinks into a forward-S shape.

As shown in Figure 1, the post-event arcade with its double row of 
footpoint brightenings begins to form after $\sim$04:36~UT.  In addition, 
in the images recorded between 04:36 and 05:12~UT, EUV brightenings 
(circled in Figure~1) may be seen at the far endpoints of the 
erupting filament; these brightenings occur as the filament threads, 
anchored well outside the filament channel itself, are jerked upward 
into a vertical orientation (see Wang et al. 2009).  From the slant 
of the post-eruption loops in the image taken at 06:00~UT, and from 
the apparent clockwise rotation of the loop orientation between 06:00 and 
07:13~UT, we conclude that the overlying arcade has right-handed helicity, 
even though the filament rotates counterclockwise.  The first post-eruption 
loops (observed at 06:00~UT) are located near the original pivot point of 
the filament rotation, where the traces approximately intersect in Figure~2, 
suggesting that the instability and rapid upward motion of the filament 
were initiated in this region.

Figure 3 shows the distribution of photospheric magnetic flux underlying 
the erupting filament.  Here, we compare a Michelson Doppler Imager 
(MDI)\footnote{See \texttt{http://soi.stanford.edu}.} line-of-sight 
magnetogram taken at 04:48~UT with the difference between the 19.5~nm images 
recorded at 04:48 and 04:36~UT.  The magnetogram, saturated at $\pm$50~G, 
represents a 5-minute average of higher-cadence data.  The circular 
filament channel evidently encloses a positive-polarity region.  
Since the northern (southern) end of the erupting filament is rooted 
in negative-polarity (positive-polarity) network flux, the axial field 
of the filament points toward the left as viewed from the 
positive-polarity region, and the filament is sinistral according to 
the chirality definition of Martin (1998).  Correspondingly, 
an H$\alpha$ image taken on September~18 at the Big Bear Solar 
Observatory (BBSO)\footnote{See \texttt{http://bbso.njit.edu}.} shows 
left-bearing barbs (Figure~4, top panel).

As indicated by the \ion{He}{2} 30.4~nm images in the left column of 
Figure~5, the filament re-forms immediately after the eruption; however, 
its northern end now bends westward rather than eastward (see also the 
H$\alpha$ image in Figure~4, bottom panel).  The MDI magnetograms 
in the right column of Figure~5 suggest that this change in the shape of 
the filament, as well as the eruption itself, may have been associated with 
the emergence of a small magnetic bipole to the west of the original 
filament channel.  The new bipole (circled in the magnetograms taken at 
06:27 and 13:19~UT) appears to have altered the shape of the photospheric 
neutral line in the vicinity of the filament.

In association with this filament event, the Large Angle and Spectrometric 
Coronagraph (LASCO) on {\it SOHO} observed a faint halo coronal mass ejection 
(CME) with an average speed of 600~km~s$^{-1}$.

\section{EVENT OF 2001 SEPTEMBER 28}
The sequence of EIT \ion{He}{2} 30.4~nm images in Figure 6, taken at 
6-hourly intervals, shows the disappearance of a high-latitude filament 
in the southern hemisphere on 2001 September~28.  In the frame recorded 
at 13:19~UT, during the eruption itself, the axis of the filament 
shows a pronounced counterclockwise rotation relative to its pre-eruption 
orientation.  This kinking into a forward-S shape is seen even more 
clearly in Figure~7, where we have superposed four successive traces 
of the filament spine taken between 19:19~UT on September~27 and 13:19~UT 
on September~28, after removing the effect of the photospheric differential 
rotation.

The eruption is shown in more detail in the sequence of \ion{Fe}{12} 19.5~nm 
running-difference images in Figure~8.  The post-event arcade begins to form 
at 15:00~UT, with the appearance of a compact bundle of highly sheared loops 
that are almost aligned with the filament channel.  From the MDI magnetogram 
(fourth panel of Figure~8), we conclude that the western end of 
this narrow bundle of reconnected loops lies on the negative-polarity side 
of the photospheric neutral line, implying right-handed helicity.  That the 
overlying arcade is indeed right-handed is confirmed by the subsequently 
forming loops (bottom panel of Figure~8), which are rooted farther 
from the PIL and are shifted clockwise relative to the earlier loops.

H$\alpha$ images recorded at BBSO on September~27 show that the 
pre-eruption filament had left-bearing barbs (see Figure~9), consistent with 
sinistral chirality.

\section{DISCUSSION}
Filaments are conventionally regarded as cool material supported inside 
the dips of a helical flux rope of a given ``handedness.''  If the 
flux rope becomes unstable, its axis rotates about the direction of ascent, 
with the rotation being clockwise (counterclockwise) if the flux rope 
has right-handed (left-handed) helicity.  Although observations of 
filament eruptions have tended to support this picture, we have here 
described two events in which the filament rotated in the opposite sense 
to that implied by the handedness of the surrounding arcade.  Both of 
these quiescent filaments were located in the southern hemisphere and 
had left-bearing barbs; their axes rotated counterclockwise, even though 
the post-event arcades were clearly right-handed.  By helicity conservation 
(Berger 1984), the source regions must have had mixed helicity, 
since the left-handed writhe acquired through the counterclockwise rotation 
could only have originated in negative helicity embodied as twist or shear 
in the pre-eruption field.

If the filament barbs are concave-downward features, as advocated by 
Martin (1998), their helicity sign or handedness would be opposite 
to that of the overlying coronal loops.  The direction of rotation 
of the erupting structure would then depend on whether the barbs 
or the overlying arcade provides the main contribution to the net helicity.  
From a preliminary survey of the entire EIT database for solar cycle~23, 
we were able to determine unambiguously the direction of rotation 
of an erupting filament in only $\sim$10 events.  Presumably because of 
the relatively low cadence of the EIT observations, all of these events 
involved non--active-region filaments, of which only the two discussed here 
rotated in the ``wrong'' sense.  In the active region events examined 
by Rust \& LaBonte (2005) and Green et al. (2007), the acquired writhe 
of the erupting filament and the helicity observed in the surrounding field 
were always of the same sign.  Thus, if the barbs and the surrounding field 
have opposite helicity signs, then the helicity of the overlying arcade 
must dominate in most eruptions, particularly those occurring in 
active regions.  That the exceptions found here involve quiescent filaments 
might be due to the fact that the latter are more likely to have 
well-developed barb systems than active region filaments.  As suggested 
by Wang (2001), barbs form when flux elements that have become connected 
to the overlying axial field diffuse across the filament channel 
and undergo flux cancellation on the other side.

That the helicity tends to be concentrated in the surrounding field 
rather than in the filament itself is consistent with recent models of 
active region fields, in which a flux rope is inserted into a potential 
configuration and the system is allowed to relax, including the effect of a 
helicity-conserving magnetic diffusion (Bobra et al. 2008; Su et al. 2009).  
The observed H$\alpha$ filament and surrounding coronal-loop structure 
are best fit with a highly sheared but weakly twisted flux rope held down 
by the overlying arcade; as in a hollow solenoid, the electric currents 
and magnetic helicity are concentrated in a semi-cylindrical shell around the 
outer edge of the flux rope, not within its interior, where the field lines 
are essentially straight and not much current flows.  Moreover, 
for stability, the amount of axial flux must be small compared with the 
total flux that holds down the filament.  Thus it is not surprising that 
it is the helicity of the surrounding field that usually determines the 
direction of rotation of an erupting active-region filament.

The event of 1999 September 20 provides support for the hypothesis that 
some filament eruptions are triggered by the emergence of flux near the 
filament channel (see, e.g., Bruzek 1952; Feynman \& Martin 1995; 
Wang \& Sheeley 1999).  In this particular case, the eruption was accompanied 
by a reconfiguration of both the underlying photospheric neutral line 
and the shape of the filament itself.  Sheeley et al. (1975) described 
a rather similar event in which the eastward hook of an H$\alpha$ filament 
disappeared and was replaced by a westward hook that pointed toward 
a region of newly emerging flux (the sense in which the filament axis rotated 
during the eruption is unknown).

We are indebted to the EIT and MDI teams for the {\it SOHO} observations, 
and to BBSO/New Jersey Institute of Technology for the H$\alpha$ images.  
We also thank A.~A. van~Ballegooijen and P.~D{\'e}moulin for 
informative discussions, and the referee for helpful comments.  This work 
was supported by NASA and the Office of Naval Research.

\clearpage
\notetoeditor{PLEASE DO NOT ROTATE THIS FIGURE!}
\begin{figure*}
\vspace{-2.5cm}
\centerline{\includegraphics[width=44pc]{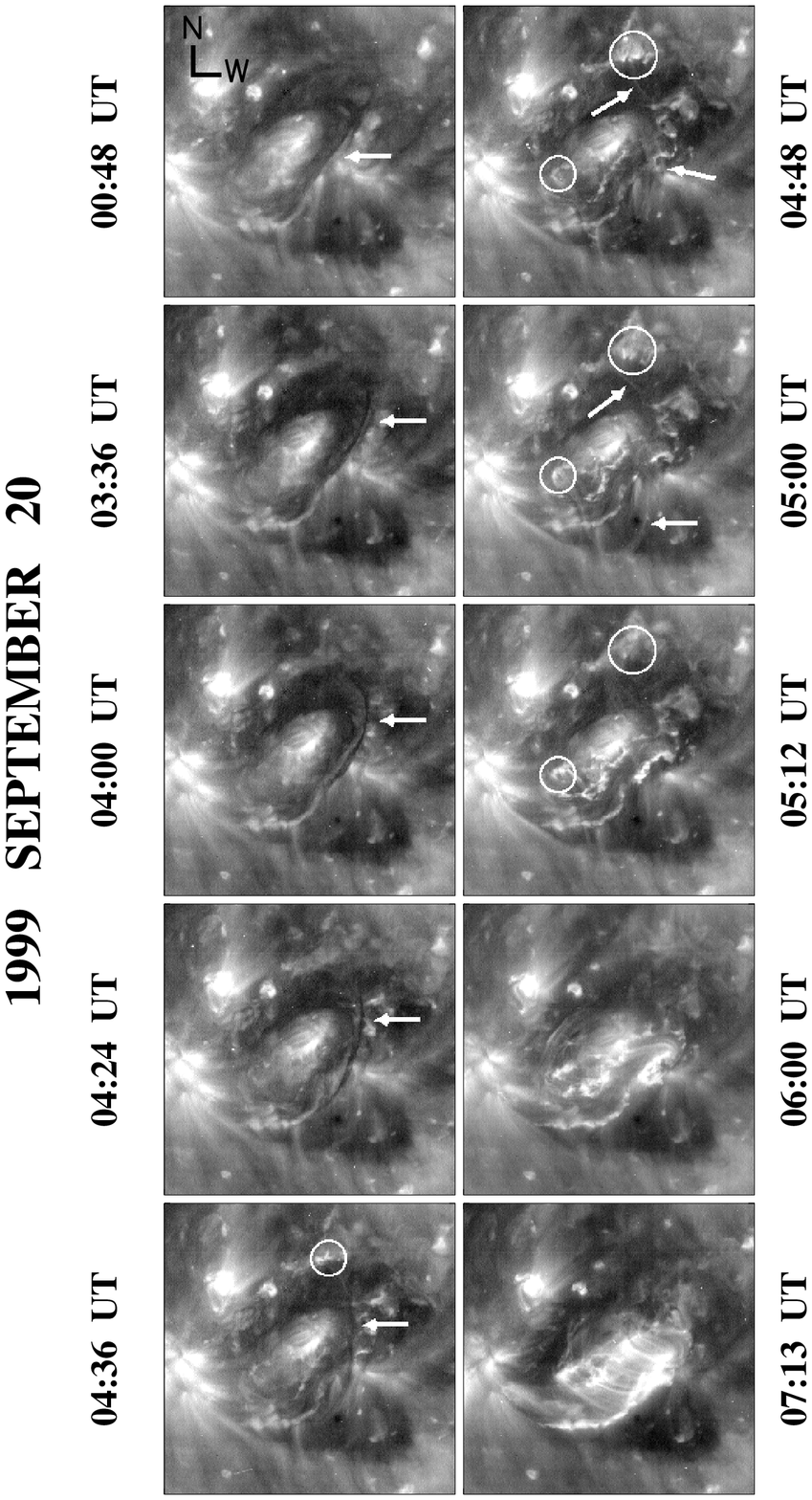}}
\vspace{-1.5cm}
\caption{Sequence of EIT \ion{Fe}{12} 19.5~nm images, showing the eruption 
of a southern-hemisphere filament on 1999 September~20.  Arrows point to 
the counterclockwise-rotating filament; brightenings marking the endpoints 
of the erupting structure are circled.  The $650^{\prime\prime}\times 
650^{\prime\prime}$ field of view is centered at longitude 1$^\circ$E, 
latitude 20$^\circ$S.  Here and in all subsequent figures, 
north is up and west is to the right (time labels run sideways).}
\end{figure*}

\clearpage
\notetoeditor{PLEASE DO NOT ROTATE THIS FIGURE!}
\begin{figure}
\vspace{-2.5cm}
\centerline{\includegraphics[width=44pc]{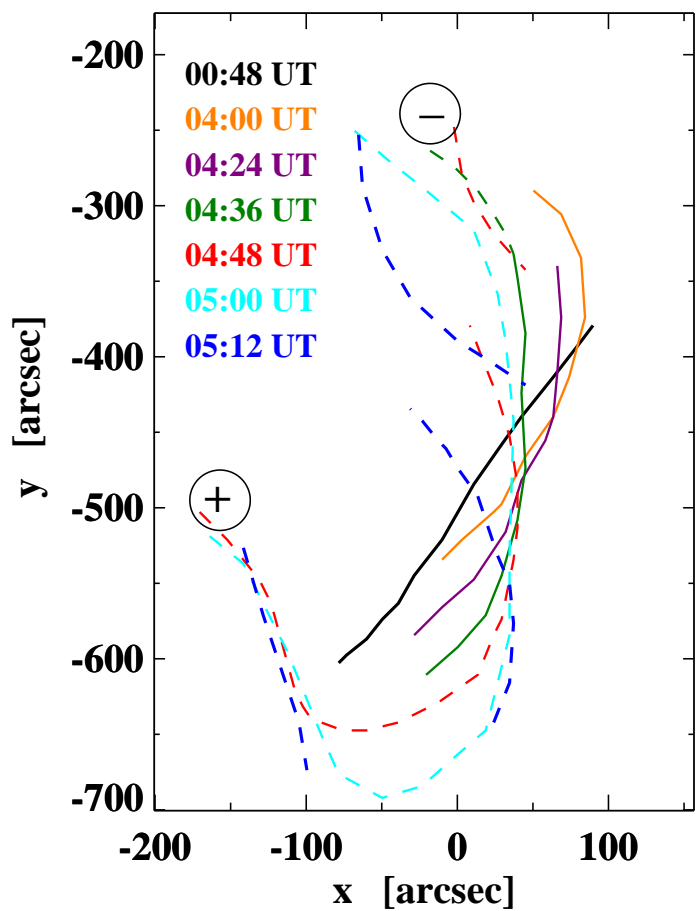}}
\vspace{0cm}
\caption{Successive traces of the dark filament spine in Figure 1 are shown 
superposed, after correcting for the longitudinal displacement (relative to 
00:48~UT, 1999 September~20) due to the average solar rotation.  
Black: 00:48~UT.  Orange: 04:00~UT.  Purple: 04:24~UT.  Green: 04:36~UT.  
Red: 04:48~UT.  Light blue: 05:00~UT.  Dark blue: 05:12~UT.  Circles with 
enclosed plus/minus signs mark the locations and polarities of the filament
endpoint brightenings (see also Figure~3).  Dashed lines indicate parts of 
the \ion{Fe}{12} 19.5~nm filament that are beginning to be seen in emission 
rather than absorption.  Note that the filament broadens and separates into 
multiple threads after 04:48~UT; the tracings at 04:48~UT and 05:12~UT 
consist of two or more disjoint pieces, because we were unable to track 
any single strand from one endpoint to the other in the \ion{Fe}{12} 19.5~nm 
images.}
\end{figure}

\clearpage
\notetoeditor{PLEASE DO NOT ROTATE THIS FIGURE!}
\begin{figure}
\vspace{-4.5cm}
\centerline{\includegraphics[width=44pc]{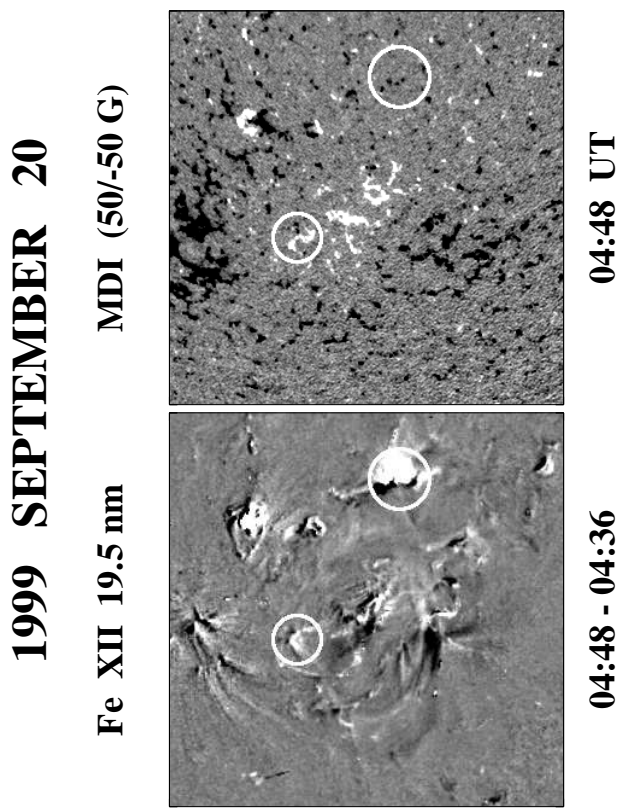}}
\vspace{-5cm}
\caption{Photospheric flux distribution underlying the filament eruption 
of 1999 September~20 (same field of view as in Figure~1).  
Top: {\it SOHO}/MDI line-of-sight magnetogram recorded at 04:48~UT 
(5-minute average of higher-cadence data).  Gray-scale levels for the 
line-of-sight field range between $B_{\rm los} < -50$~G (black) and 
$B_{\rm los} > +50$~G (white).  Bottom: \ion{Fe}{12} 19.5~nm 
running-difference image taken at 04:48~UT.  The locations of the 
brightenings at the endpoints of the erupted filament are circled in 
both the MDI and the EIT image.  As viewed from the positive-polarity region 
inside the circular filament channel, the axial field points to the left, 
so that the filament is sinistral.}
\end{figure}

\clearpage
\notetoeditor{PLEASE DO NOT ROTATE THIS FIGURE!}
\begin{figure}
\vspace{-6cm}
\centerline{\includegraphics[width=44pc]{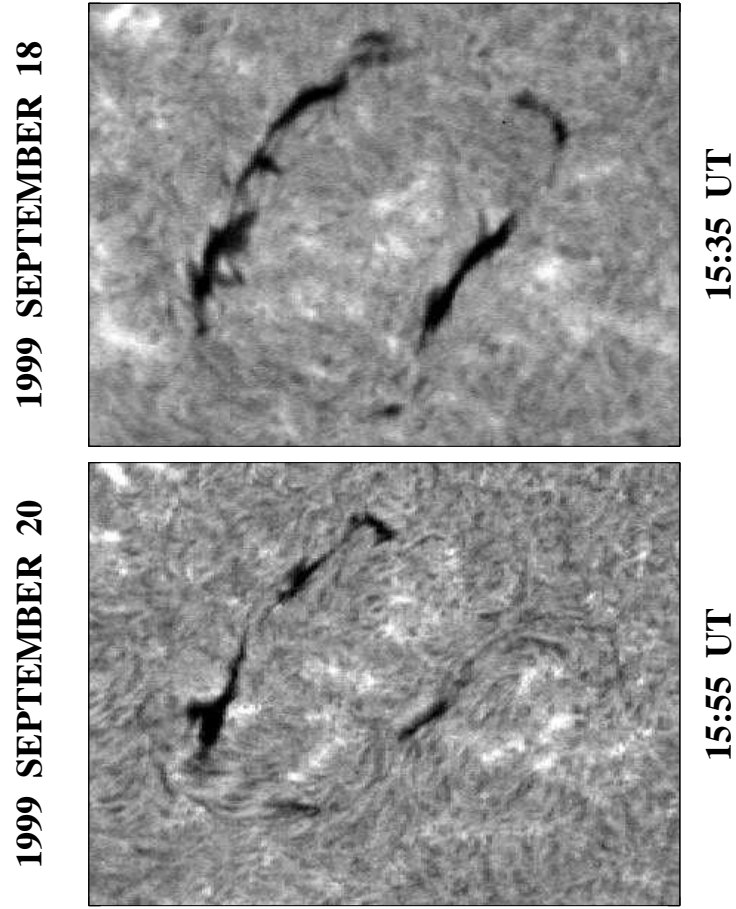}}
\vspace{-4cm}
\caption{BBSO H$\alpha$ images recorded on 1999 September 18 (top) 
and 1999 September~20 (bottom).  The left-bearing barbs of this 
southern-hemisphere filament are clearly seen in the pre-eruption image.  
After the filament erupted early on September~20, the re-formed structure 
hooks westward instead of eastward (see also Figure~5).}
\end{figure}

\clearpage
\notetoeditor{PLEASE DO NOT ROTATE THIS FIGURE!}
\begin{figure*}
\vspace{-3cm}
\centerline{\includegraphics[width=44pc]{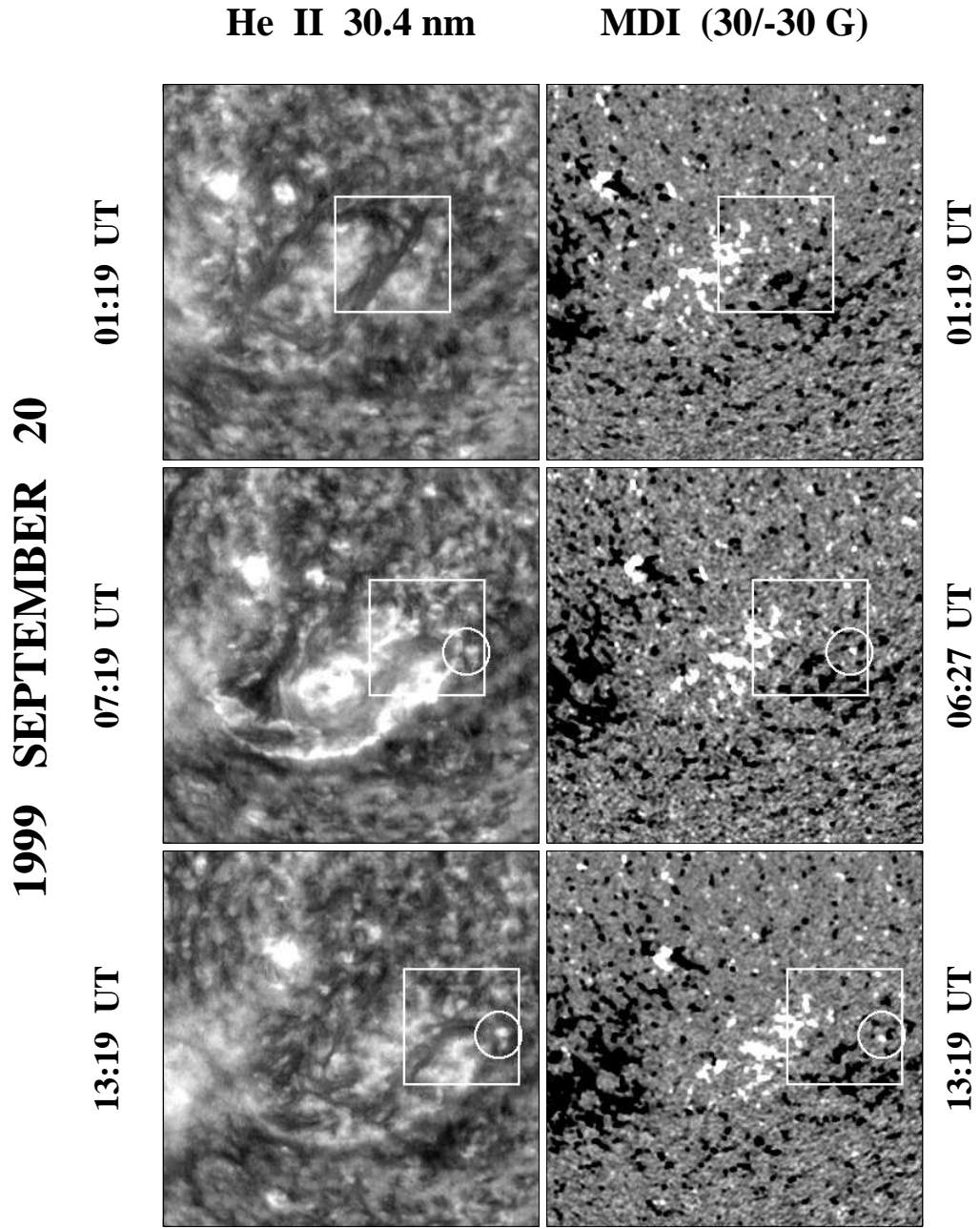}}
\vspace{-2.5cm}
\caption{The event of 1999 September 20 coincided with a change in the 
shape of the filament and the underlying photospheric neutral line.  
Left panels: sequence of \ion{He}{2} 30.4~nm images taken at 
01:19, 07:19, and 13:19~UT.  Right panels: corresponding MDI 
line-of-sight magnetograms (smoothed 5-minute averages, saturated at 
$\pm$30~G).  A small magnetic bipole (circled) emerged to the west 
of the original filament channel just before the eruption.  As in 
Figure~1, the field of view has dimensions $650^{\prime\prime}\times 
650^{\prime\prime}$ and is centered at (1$^\circ$E, 20$^\circ$S).}
\end{figure*}

\clearpage
\notetoeditor{PLEASE DO NOT ROTATE THIS FIGURE!}
\begin{figure*}
\vspace{-3cm}
\centerline{\includegraphics[width=44pc]{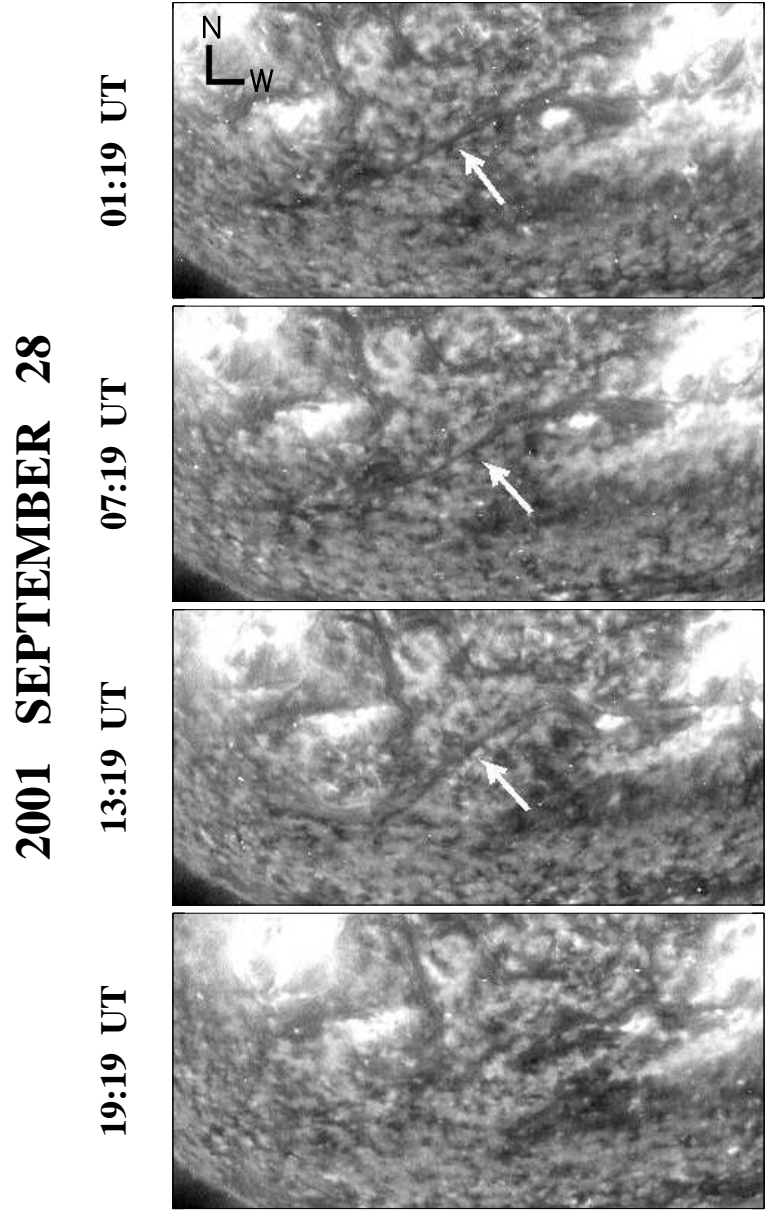}}
\vspace{-2.5cm}
\caption{Sequence of EIT \ion{He}{2} 30.4~nm images taken at 6-hourly 
intervals, showing the disappearance of a southern-hemisphere filament 
on 2001 September~28.  The filament axis has clearly undergone a 
counterclockwise twist at 13:19~UT, in the midst of the eruption itself.  
The $1040^{\prime\prime}\times 520^{\prime\prime}$ field of view is 
centered at (5$^\circ$E, 32$^\circ$S).}
\end{figure*}

\clearpage
\notetoeditor{PLEASE DO NOT ROTATE THIS FIGURE!}
\begin{figure}
\vspace{-2.5cm}
\centerline{\includegraphics[width=36pc]{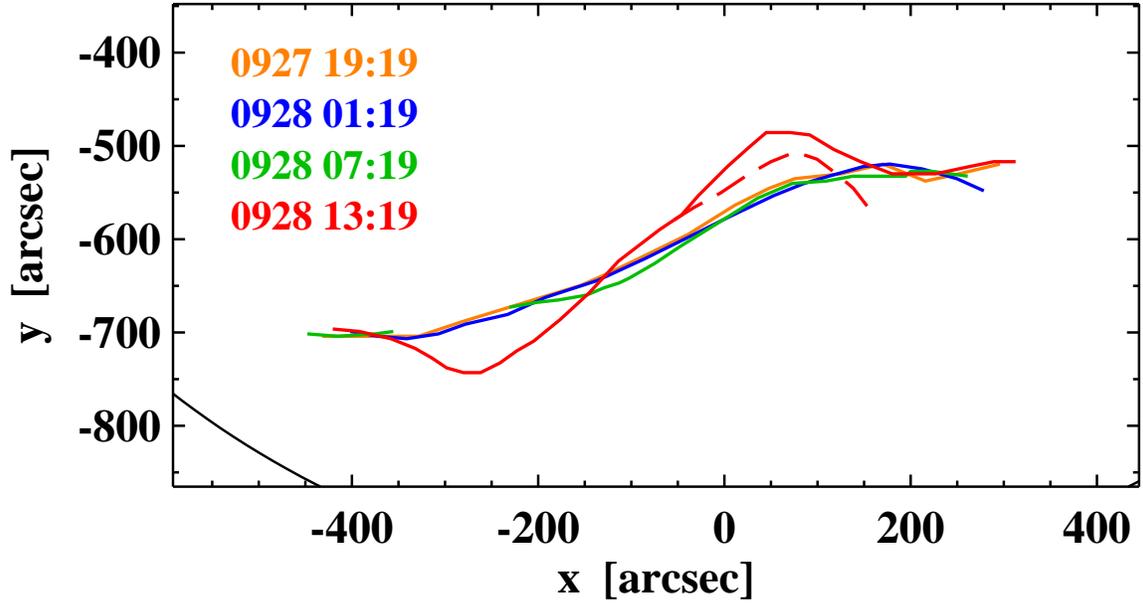}}
\vspace{0cm}
\caption{Successive traces of the filament spine in Figure 6 are shown 
superposed, after correcting for the longitudinal displacement (relative to 
13:19~UT, 2001 September~28) due to the photospheric differential rotation.  
Orange: 19:19~UT (September~27).  Blue: 01:19~UT (September~28).  
Green: 07:19~UT (September~28).  Red: 13:19~UT (September~28).  
Field of view is the same as in Figure~6.  At 13:19~UT, the western end 
of the filament splits into two branches; the southern branch is indicated 
by dashed lines.}
\end{figure}

\clearpage
\notetoeditor{PLEASE DO NOT ROTATE THIS FIGURE!}
\begin{figure*}
\vspace{-3cm}
\centerline{\includegraphics[width=44pc]{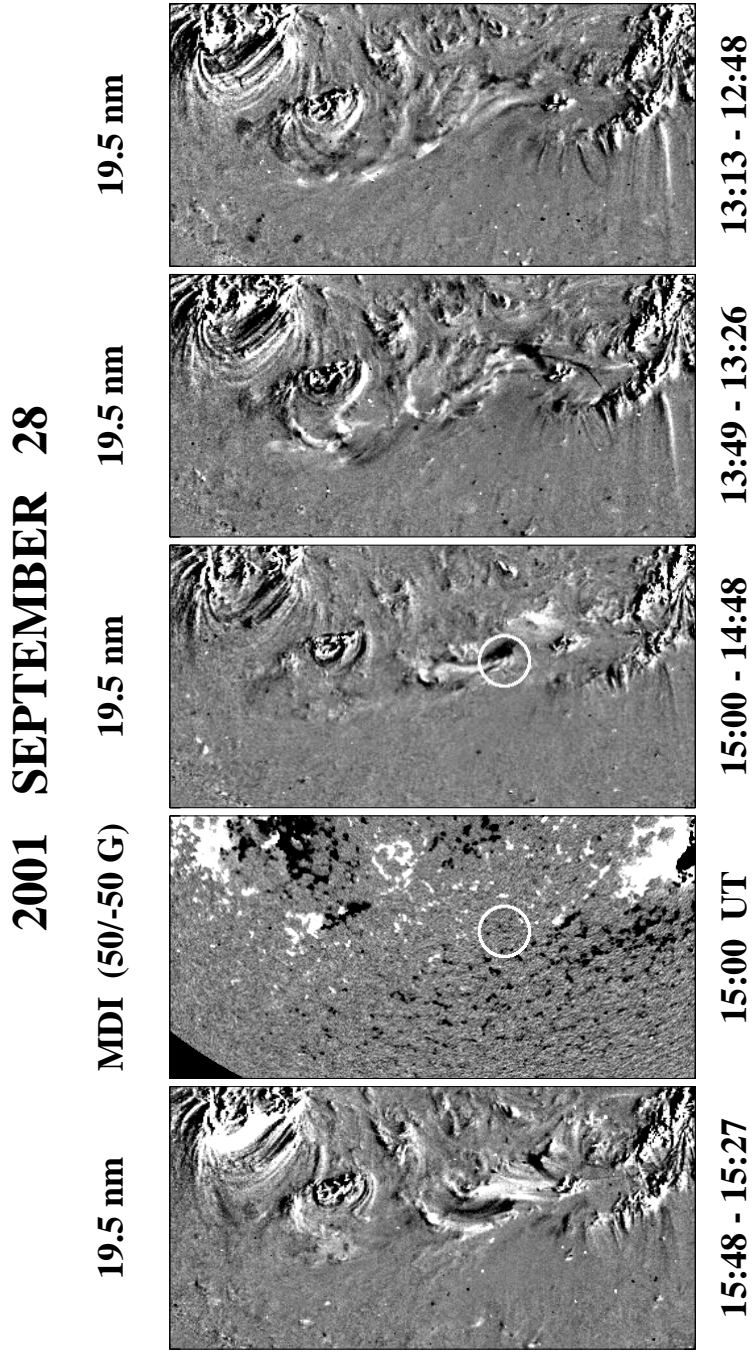}}
\vspace{-2cm}
\caption{A closer look at the 2001 September~28 filament eruption 
(same field of view as in Figure~6).  Top three panels: 
\ion{Fe}{12} 19.5~nm running-difference images taken at 13:13, 13:49, 
and 15:00~UT.  Fourth panel: MDI line-of-sight magnetogram 
recorded at 15:00~UT (5-minute average of higher-cadence data).  
Bottom panel: 19.5~nm running-difference image taken at 15:48~UT, 
where the post-event arcade is seen to have right-handed helicity.  
The first compact bundle of reconnected loops, appearing at 15:00~UT, 
is highly sheared and almost aligned with the filament channel; 
its westernmost end (circled) lies on the negative-polarity side of the PIL, 
again indicating right-handed helicity.}
\end{figure*}

\clearpage
\notetoeditor{PLEASE DO NOT ROTATE THIS FIGURE!}
\begin{figure}
\vspace{-8cm}
\centerline{\includegraphics[width=44pc]{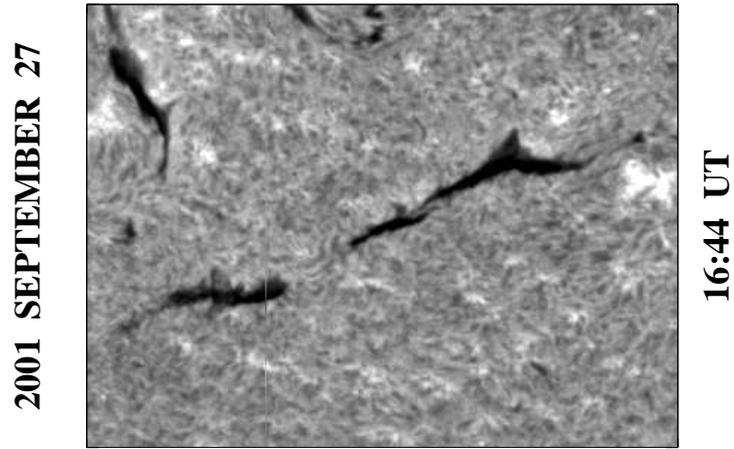}}
\vspace{-7cm}
\caption{BBSO H$\alpha$ image recorded on 2001 September 27, showing 
the left-bearing barbs of the southern-hemisphere filament.}
\end{figure}

\end{document}